\documentclass[]{aipproc}
\layoutstyle{6x9}
\begin{document}

\title[The
$D^0 \to K^0 \bar K^0$ decay beyond factorization]
{The
$D^0\to K^0\bar K^0$ decay beyond factorization
\footnote{Talk given by S. Fajfer}}

\author{J. O. Eeg}{
 address={Dept. of Physics, Univ. of Oslo, N-0316 Oslo, Norway},
}

\iftrue
\author{S. Fajfer}{
  address={Physics Department, University of Ljubljana, Jadranska 19,
and Institut Jo\v{z}ef Stefan, Jamova 39, SI-1000 Ljubljana, 
Slovenia},
}
\author{J. Zupan}{
  address={Institut Jo\v{z}ef Stefan, Jamova 39, SI-1000 Ljubljana,
Slovenia}
}
\fi

\copyrightyear  {2001}

\begin{abstract}
The decay mode $D^0 \rightarrow K^0 \bar K^0$ has no
factorizable contribution.  We calculate
the nonfactorizable  chiral
loop contributions of order ${\cal O}(p^3)$ and then we use a
heavy-light
type chiral quark model to calculate nonfactorizable  tree level terms,
 also of order ${\cal O}(p^3)$, proportional to the gluon condensate.
Calculated chiral  loops and the gluon condensate
contributions
 are of the same order of magnitude as the experimental amplitude.

\end{abstract}

\date{\today}

\maketitle


For nonleptonic decays of  $D$ mesons \cite{WSB} - \cite{BLMPS} as well
as  for
$K$'s and $B$'s,
the so called {\em factorization} hypothesis has been commonly used.
The factorization hypothesis are known to fail badly for nonleptonic
$K$ decays \cite{K-fact,BEF}.
On the
other hand, there are  certain heavy hadron weak decays where
factorization might apply.
Recently,  the understanding of factorization  for
exclusive nonleptonic decays of $B$ mesons in terms of QCD in the
 heavy quark limit has
been considerably improved  \cite{BBNS}. Following \cite{EFZ}
we discuss nonfactorizable terms for $D$ decays,
in particular for the decay mode $D^0\to K^0\bar{K^0}$.
 In $D^0\to K^0\bar{K^0}$,  factorization   misses
completely, predicting a vanishing branching ratio, in contrast with the
experimental situation.
To see this, note that at tree level the $D^0\to K^0\bar{K^0}$ decay
might
occur due to two annihilation diagrams \cite{WSB} which could
potentially
create the $K^0 \bar K^0$ state. However, they cancel each other by the
GIM mechanism. Moreover, in factorization limit, the amplitude is
proportional to
\begin{equation}
\langle  K^0 \bar K^0| V_\mu |0\rangle  \langle 0 | A^\mu | D^0\rangle
\simeq (p_{K^0} - p_{\bar K^0})_\mu \,  f_D p_{D}^\mu= 0 \; .
\label{eq-9}
\end{equation}
In many of the studies (e.g.
\cite{XYP,DDL,HL,GPW,Z}) this decay has been understood as a
result of final state interactions (FSI) e.g. \cite{XYP}.
 A recent investigation of the $D^0\to K^0\bar {K^0}$ decay mode
performed in
\cite{DDL} has focused on the  $s$ channel  and the $t$ channel one
particle exchange contributions.

On the other hand it
is well known that  factorization does not work in  nonleptonic $K$
decays. Among many approaches the Chiral Quark Model
 ($\chi$QM) \cite{pider} was shown to be able to accommodate the
intriguing  $\Delta I = 1/2$ rule  in
$K \to \pi \pi$ decays,  as well
as CP violating parameters, by systematic involvement of the soft
gluon emission forming gluon condensates and chiral
loops at ${\cal O} (p^4)$ order \cite{BEF}.
In the $\chi$QM \cite{pider} 
the light quarks ($u, d, s$) couple to the would-be Goldstone octet
mesons ($K, \pi, \eta$) in a chiral invariant way, such that all effects
are in principle calculable in terms of physical quantities and a few
model dependent parameters, namely the quark condensate, the gluon
condensate
and the constituent quark mass \cite{BEF,pider,epb}.

In the case of $D$ meson decays one
has to extend the ideas of the $\chi$QM to the sector involving a
heavy quark ($c$) using the chiral symmetry of light degrees of 
freedom
as well as heavy quark symmetry and
Heavy Quark Effective Field Theory (HQEFT).
  Such ideas have already been presented
in previous papers \cite{barhi,effr,itCQM} and lead to the 
formulation
of  Heavy-Light Chiral Quark Models (HL$\chi$QM).
In our formulation of the  HL$\chi$QM Lagrangian, an unknown 
coupling
constant
appears in  the term  that couples  the heavy meson to a heavy and
 a light quark \cite{EFZ}.
Our strategy is  to relate  expressions involving this coupling to
physical
 quantities, as it is done within the $\chi$QM \cite{BEF}. We perform
the bosonization by integrating out the light and heavy quarks and
obtain
a heavy quark symmetric chiral Lagrangian involving light and heavy
mesons \cite{itchpt,wise,stewart}.

Because the ${\cal O}(p)$ (factorizable) contribution is zero
as seen in Eq. \eqref{eq-9},
we  approach to the $D^0 \to K^0 \bar K^0$
decay by calculating 
systematically ${\cal O}(p^3)$ contributions. We do this by including first
 the nonfactorizable
 contributions coming from the chiral loops. These are based on the weak
 Lagrangian
corresponding  to the factorizable  ${\cal O}(p)$ terms for
$D^0 \rightarrow \pi^+ \pi^-$ and $D^0 \rightarrow K^+ K^-$.
Second, we consider the  gluon condensate contributions, also of
${\cal O}(p^3)$ within the $\chi$QM and HL$\chi$QM framework.
The energy release in
$D \rightarrow K \bar{K}$ is $p =788$
MeV and hence $p/\Lambda_\chi$ (for $\Lambda_\chi\geq 1$ GeV),
is close to unity.
The next to leading ${\cal O}(p^5)$ terms might
be almost
of the same order of magnitude compared to our ${\cal O}(p^3)$ terms.
However, we expect a weak suppression of the order
$p^2/\Lambda_\chi^2$.
On the other hand, the inclusion of
${\cal O}(p^5)$ order in this framework is not straightforward.
Before doing loop calculations at that order,
one has to find  a reliable framework to include
 light resonances like  $\rho$, $K^*$, $a_0(980)$, $f_0(975)$ etc.
 The poorly
known scalar resonances  would introduce a rather
large uncertainty \cite{DDL}. Right now, a 
consistent calculation of this or higher orders does not seem to be
possible.

Note that we have  omitted $1/m_Q$ terms in the framework of HQEFT.

The effective weak Lagrangian at quark level
relevant for  $D \rightarrow \pi \pi, K \bar{K}$ is
\begin{equation}
 {\cal L}_{W}= \widetilde{G} \left[ c_A \, (Q_A -Q_C) \;
+ c_B \, (Q_B^{(s)} - Q_B^{(d)}) \right] \; ,
 \label{Lquark}
\end{equation}
where
$\widetilde{G} =  -   \, 2 \sqrt{2} G_F V_{us}\,V^*_{cs}$, and
\begin{eqnarray}
Q_A  =   ( \overline{s}_L \gamma^\mu  c_L )  \; \,
           ( \overline{u}_L \gamma_\mu  s_L )
\; \;  , \,
Q_{C}  =  ( \overline{d}_L \gamma^\mu c_L ) \; \,
          ( \overline{u}_L  \gamma_\mu d_L )
\, , \nonumber  \\
Q_B^{(q)}  =   \, ( \overline{u}_L \gamma^\mu c_L ) \; \,
           ( \overline{q}_L \gamma_\mu q_L )
\; \; , \qquad
 (q  \;  = \; s,d)
\, ,
\label{QA-QC}
\end{eqnarray}
are quark operators.

Using Fierz transformations \cite{EFZ}
one obtains operators
$Q_A   = Q_B^{(s)}/N_c  +  R_B^{(s)}$,
$Q_C  =  Q_B^{(d)} /N_c + R_B^{(d)} $, 
$Q_B^{(s)} =  Q_A/N_c  + R_A$
and
$Q_B^{(d)}=  Q_C/N_c  + R_C $, where the $R$'s  correspond to color
exchange between two currents and
are genuinely nonfactorizable:
$R_A = 2 ( \overline{s}_L \, \gamma^\mu \, t^a \, c_L \, )
( \overline{u}_L \, t^a \, \gamma_\mu \, s_L \,) $,
$R_C = 2  ( \overline{d}_L \, \gamma^\mu \, t^a \, c_L \, )
( \overline{u}_L \, t^a \, \gamma_\mu \, d_L \,) $,
$ R_B (q) = 2 ( \overline{u}_L \, \gamma^\mu \, t^a \, c_L \,)(
\overline{q}_L
\, t^a \, \gamma_\mu \, q_L \,)$ ,
 $(q  \;  = \; s,d)$. The operators can be written in terms of currents
\cite{EFZ}.
The factorization approach  amounts to writing the currents
 in terms of hadron (in our case meson)
fields only, so that the operator $Q_B^{(s)} - Q_B^{(d)}$  in the
left equation is equal to the product of two
meson currents. The color currents  are then zero if
hadronized (mesons are color singlet objects).
There is also a replacement of the Wilson coefficients in the hadronized
effective
weak Lagrangian $c_{A,B}\to c_{A,B}(1+1/N_c)$.
Combining heavy quark symmetry and chiral symmetry of the light sector,
we can obtain the weak chiral Lagrangian for nonleptonic
$D$ meson decays due to
factorizable terms. Then we can first use this to
calculate  nonfactorizable contributions due to chiral loops.
 Second, we can  calculate
 the color currents' contribution  using the
gluon  condensate within
the framework of the HL$\chi$QM.

Treating the light pseudoscalar mesons as pseudo-Goldstone bosons one
obtains
the  usual ${\cal O}(p^2)$ chiral Lagrangian \cite{EFZ} and from
this Lagrangian,
 we can deduce the
 light weak current  to  ${\cal O}(p)$.

In the heavy meson sector interacting with light mesons we have
used
the lowest order ${\cal O}(p)$  chiral Lagrangian (for details see
\cite{EFZ}). Using symmetry arguments, 
the heavy-light weak current is bosonized to  ${\cal O}(p^0)$,  
with the unknown coupling $\alpha_H$ related to the physical decay
 constant $f_D$.


In the factorization limit there are no contributions
to $D^0\to K^0 \bar{K^0}$ at tree level. The observation of a partial
decay width  $B(D\to K^0\bar{K^0})=(6.5\pm 1.8)\times 10^{-4}$ on the
other hand implies that we can expect sizable contributions at the one
loop level. Calculations to one loop  in the framework of combined
chiral
perturbation theory and HQEFT involves  a
construction of the most general effective Lagrangian that has the
correct
symmetry properties in order to make the renormalization work.
Our calculations were done in the strict $\overline{MS}$
renormalization scheme.

Writing down the most general one loop graphs with two outgoing
Goldstone
bosons ($K^0$ and $\bar{K^0}$)  one arrives at 26 Feynman
diagrams. 
The expressions for nonzero amplitudes corresponding to the graphs on
Fig. 1
are given in \cite{EFZ}.

\begin{figure}[!t]
\resizebox{.65\columnwidth}{!}
{\includegraphics{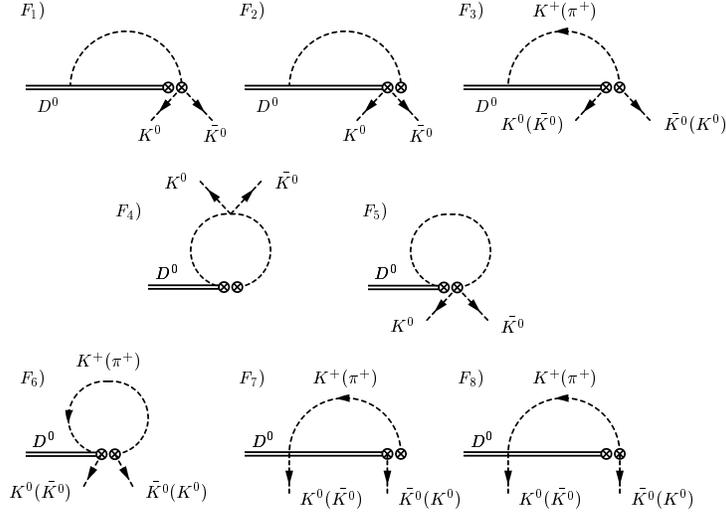}}
\caption{The diagrams which give nonzero amplitudes. }\label{fig-5}
\end{figure}

The partial decay
width
for the decay $D^0 \to K^0\bar{K^0}$ is then
\begin{equation}
\Gamma_{D^0\to
K^0\bar{K^0}}=\frac{1}{2\pi}\frac{G_F^2}{8m_D}
c_A^2  |V_{us} \, V^*_{cs}|^2
\frac{|F|^2}{(8\pi^2)^2}
p \; ,\label{eq-2}
\end{equation}where
$F=\sum_n F_n$ is the sum of the amplitudes corresponding to the graphs
on Fig.1 and $p$ is the $K^0(\bar{K^0})$ three - momentum in the
$D^0$ rest frame.

In  numerical calculation  we use the values of $\alpha_H$, $g$ and
$f$ ($f$ is related to the $\pi$ meson decay constant) 
obtained within the same framework  \cite{itchpt,stewart,
Grinstein-92,
Grinstein-94,BG}.
The coupling $g$ is extracted from existing
experimental data  on $D^*\to D\pi$ and $D^*\to
D\gamma$ decays. The analysis  in \cite{stewart} includes chiral
corrections at
one
loop order and  yields
$g=0.27^{+0.04+0.05}_{-0.02-0.02}$, leaving
the sign undetermined.
Recently CLEO Collaboration has measured the  $D^{*+}$ decay
width \cite{cleo-01}. By combining this result with existing
data on $D^*$ decay widths
\cite{PDG-00}, we obtain value $g=0.57 \pm 0.08 $.  We present results
for $g=0.27$ and  $g=0.57$. The larger  value seems to be in better
agreement
with the results coming from different approaches listed in
\cite{itchpt}. We put everywhere  $\mu = 1$ $\rm{GeV} \simeq \Lambda_\chi$.

For the Wilson coefficients $c_{A,B}$ of (\ref{Lquark})
 we use $c_A=1.10\pm 0.05$ and $c_B=-0.06\pm 0.12$
\cite{buras}, calculated   at the scale $\mu=1$ $
\rm{GeV}$ with the number of colors $N_c=3$.
Due to the
 suppression of  $c_B$ in comparison with $c_A$, we do not include
 terms proportional to
$c_B$. 

\begin{table} [h]
\begin{tabular}{lcc} \hline
&\tablehead{1}{c}{b}{${\cal M}_i[\times 10 ^{-7}{\rm GeV}] $ $(
g=0.27)$}
&\tablehead{1}{c}{b}{${\cal 
M}_i[\times 10 ^{-7}{\rm GeV}] $ $(g=0.57)$}\\
\hline 
${\cal M}_1$ & -0.42 & -0.82 \\
${\cal M}_2$ & -0.31 & -0.62\\ 
${\cal M}_3$ & -0.62 & -1.23 \\ 
${\cal M}_4$ & 0.75 -2.54 i & 0.70 -2.37 i \\ 
${\cal M}_5$ &-0.81 & -0.76 \\ 
${\cal M}_6$ &-0.61 & -0.57 \\ 
${\cal M}_{7}$ &-0.99 & -0.92 \\ 
${\cal M}_{8}$ & 0.91 & 0.85\\ 
\hline
$\sum_i {\cal M}_i$& -2.11 -2.54 i &  -3.37 -2.37 i\\ \hline
\end{tabular}
\caption{{Table of the one chiral loop  amplitudes (see
Fig. 1), 
where ${\cal M}=\sum_n {\cal M}_n$. 
 The second column shows the amplitudes calculated using
$g=0.27$ 
while  the third column
amplitudes have been calculated using $g=0.57$.
}}
\label{tab-1}
\end{table}
The imaginary
part of the amplitude comes from the $F_4$ graph, when  the  $\pi$'s
or the  $K$'s  in the loops are on-shell. All other
 graphs contribute only to the
real part of the amplitude.
The imaginary part of the amplitude is scale and scheme independent
within chiral perturbation theory.
This amplitude is also obtained from unitarity, and is
valid beyond the chiral loop expansion.

\begin{figure}
 \resizebox{.2\columnwidth}{!}{\includegraphics{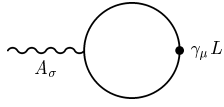}}
\caption{Feynman diagram for bosonization of left-handed current
to order ${\cal O}(p)$}\label{fig:va}
\end{figure}

\begin{figure}
 \resizebox{.2\columnwidth}{!}{\includegraphics{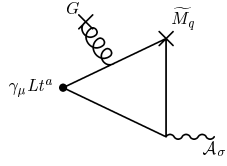}}
\caption{Diagram for bosonization of the color current to ${\cal
O}(p^3)$}
\label{fig:colcur}
\end{figure}

\begin{figure}
 \resizebox{.4\columnwidth}{!}{\includegraphics{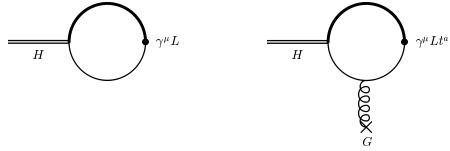}}
\caption{Diagrams representing bosonization of heavy-light weak current.
The boldface line represents the heavy quark, the solid line the light
quark.}
\label{fig:heavylight}
\end{figure}


In the effective weak Lagrangian
there are, after Fierz transformations, terms that
involve
 color currents. As mesons are color
 singlet objects, the product of color currents does not contribute at
meson level in the factorization limit. However, at quark level they do
 contribute through  the gluon
 condensate. In order to
 estimate this contribution we have to establish the connection between
the underlying quark-gluon dynamics and the meson level picture. This is
 done through the use of a Heavy-Light Chiral Quark Model
(HL$\chi$QM).
Our starting point is the  Lagrangian containing both quark
 and meson fields \cite{EFZ,barhi,effr,itCQM,AJ}.
After bosonizing  the light weak current at the order ${\cal O}(p)$
and  ${\cal O}(p^3)$ (see Fig. 2 and Fig. 3) as in \cite{EFZ} one should
bosonize
the heavy - light weak current integrating out quark fields as
presented
on Fig. 4. The product of two external gluon fields ($G$ 
in Fig. 3 and Fig. 4 ) is interpreted as the gluon condensate 
($\langle G^2\rangle$) \cite{EFZ}.  This contribution is of the 
 ${\cal O}(p^3)$ order. 
 In the language of chiral 
perturbation theory, the divergent part of the counterterm
has the Lorentz and flavor structure of the effective Lagrangian 
given in eq. (47) in \cite{EFZ}. 
 
By taking into account various
relations of the loop integrals we determine
 \begin{equation}
{\cal{M}}(D^0 \rightarrow K^0 \bar{K^0})_{\langle G^2\rangle }
\; \simeq \;  0.43 \times 10^{-7}  {\rm GeV} \; ;
\label{ampNum}
\end{equation}
which is of the same order of magnitude as the chiral loop
contributions
in Table 1.
Adding both the  chiral loops  and the gluon
condensate \eqref{ampNum} contributions,  we obtain the total amplitude
 to ${\cal
O}(p^3)$
\begin{eqnarray}
g=0.27 ; \enspace &
{\cal M}_{{\rm Th}}=(-1.68 - 2.54 \, i)\times 10^{-7}{{\rm GeV}} \;
\\
g=0.57 ;\enspace &
{\cal M}_{{\rm Th}}=(-2.94 -2.37 \, i)\times 10^{-7}{{\rm GeV}} \; .
\end{eqnarray}
or in terms of branching ratio
\begin{eqnarray}
g=0.27 ; \enspace &
B(D^0\to K^0\bar{K^0})_{{\rm Th}}=(4.2\pm 1.4)\times 10^{-4} \; \\
g=0.57 ; \enspace &
B(D^0\to K^0\bar{K^0})_{{\rm Th}}=(6.5\pm 1.7)\times 10^{-4}
\end{eqnarray}
where the estimated  uncertainties reflect the  uncertainties
 in the rest of input parameters. These results should be compared with
experimental data  \cite{PDG-00} $B(D^0 \to K^0
\bar K^0)
= (6.5\pm 1.8)\times 10^{-4}$ .

We can summarize that we  indicate the leading
 nonfactorizable
contributions to $D^0 \to K^0 \bar K^0$. Even though the use of chiral
perturbation theory in this decay mode could be questioned,
the calculated chiral loops
can be considered as  part of the final state
interactions.  In the treatment of the final state interactions
the light pseudoscalar meson exchanges have to  be
present  due to unitarity.
Although, the next to leading ${\cal O}(p^5)$ order terms
might give sizable contributions
to this decay, we have demonstrated that contributions due to
the chiral loops and gluon
condensates are of the same order of magnitude
as the amplitude extracted from the experimental result.



\end{document}